\documentstyle[epsf,rotate,preprint,prb,aps]{revtex}
\def\inseps#1#2{\def\epsfsize##1##2{#2##1} \centerline{\epsfbox{#1}}}
\input{mssymb.tex}

\def\be{\begin{equation}}
\def\th0{\theta_0}

\def\bea{\begin{eqnarray}}
\def\ee{\end{equation}}
\def\eea{\end{eqnarray}}

\begin{document}

\title{Phase diagram of magnetic polymers}

\author{T. Garel, H. Orland}
\address{Service de Physique Th\'eorique,\\ CEA Saclay,
91191 Gif-sur-Yvette Cedex,\\ 
France\\}
\author{E. Orlandini}
\address{Istituto Nazionale per la Fisica della Materia (INFM) and
Dipartimento di Fisica dell'Universita' di Padova, via Marzolo 8, 35131 Padova, Italy}
\date{\today}
\maketitle

\begin{abstract}
We consider polymers made of magnetic monomers (Ising or
Heisenberg-like) in a good solvent. These polymers are modeled as
self-avoiding walks on a cubic lattice, and the ferromagnetic
interaction between the spins carried by the monomers is short-ranged
in space. At low temperature, these polymers undergo a magnetic
induced first order collapse transition, that we study at the mean
field level. Contrasting with an ordinary $\Theta$ point, there is a
strong jump in the polymer density, as well as in its
magnetization. In the presence of a magnetic field, the collapse
temperature increases, while the discontinuities decrease. Beyond a
multicritical point, the transition becomes second  order and
$\Theta$-like. Monte Carlo simulations for the Ising case are in
qualitative agreement with these results. 
\end{abstract}
\vskip 3mm
\noindent\mbox{Submitted for publication to: Eur. Phys. J. B} \hfill %
\mbox{Saclay, SPhT/99-008}\newline
\smallskip
\noindent \mbox{PACS: 36.20.Ey; 64.70.-p; 75.50.-y}

\newpage

\section{Introduction}
Under various solvent conditions, a polymer chain can be either
swollen or collapsed. In a bad solvent, a phase transition between these
two states can occur as temperature is varied \cite{Flory}. This is
the so-called $\Theta$ point, 
the tricritical nature of which has been demonstrated by de Gennes
\cite{deGennes}. 
The effective monomer-monomer attraction results from 
tracing out the solvent degrees of freedom. At the $\Theta$ point,
the second virial coefficient of the polymer vanishes.

In this paper, we consider a different mechanism that also yields attractive
monomer-monomer interactions. The model we study consists of a polymer
chain, where each monomer carries a spin ${S}$ in an external magnetic 
field
; these spins interact with each other
with a short-ranged interaction. 
To be specific, we will consider a
self-avoiding walk (SAW) of length $N$, on a $d$ dimensional cubic 
lattice, with a
nearest-neighbor ferromagnetic 
interaction (on the lattice) between the spins of the monomers. 

Several models involving both polymeric and magnetic-like degrees of
freedom have been introduced in very different contexts. 
A similar, but somewhat 
more complicated model, was studied to describe secondary structure
formation in proteins \cite{PGO}: there, the dominant 
interactions between the monomers are  of (electric) dipolar nature.
A 
Potts model on a SAW was studied as a description
of vulcanization in \cite{Co_Da}; a similar model with quenched
disorder was studied in the context of secondary structure formation
in proteins \cite{Ar_Sha}. 

Ising models have also been studied on a
fixed SAW geometry, yielding results quite different from those
presented here \cite{Cha_Ma_Sti,Aer_Van}. From an experimental point
of view, organic polymeric magnets have recently become of interest 
\cite{Mi,Ep_Mi}. However, there has been very little interest in their 
conformational changes. 

In the following, we will
specialize to the Ising case, and quote some results for the Heisenberg 
case.
The partition function of the system reads:
\be
\label{Z}
Z= \sum_{\rm SAW} \sum_{S_i=\pm 1} \exp\left({{\beta J \over 2} \sum_{i \ne j}
S_i \Delta_{r_i r_j} S_j + \beta h \sum_i S_i}\right)
\ee
where $J$ is the exchange energy, $\beta={1 \over T}$ the inverse
temperature, and $h$ the external magnetic field. The spatial
position of monomer $i$ with spin $S_i$ is 
$r_i$. The symbol $\Delta_{r r'}$ is 1 if $\{r,r'\}$ are
nearest-neighbor on the lattice and $0$ otherwise.
The sums run over all possible SAW and all spin configurations.

This model will be studied along three lines. In section
\ref{bounds}, we derive upper bounds for the free energy of the model 
(in zero magnetic field) which suggest a first order transition
between a swollen paramagnetic and a collapsed magnetized phase. In
section \ref{mean_field}, we derive a mean-field theory for the general
model. We show that indeed for low fields, there is a such a line of
first-order transitions. At higher fields, this transition becomes
continuous. The two regimes are separated by a multicritical $G$
point. At this special point, both the second and the third virial
coefficient vanish. In section \ref{MC}, these predictions are tested
against Monte Carlo simulations in $d=3$ dimensions. The numerical results are
consistent with the theoretical phase diagram; at low field (where the
transition is strongly first order), the agreement is even quantitative.

\section{Free energy bounds}
\label{bounds}
The physics of the model can be described according to the following
simple picture:
\begin{enumerate}
\item{} At high temperature, since entropy dominates, the chain is swollen.
As a result, the number of nearest-neighbor contacts is small, 
and 
from a magnetic point of view, the system is equivalent to a one dimensional
Ising model. This simple picture can be expressed through the following
inequality:
\be
\label{bound1}
Z  \ge Z_{\rm SAW}\  Z_1 (h)
\ee
where $Z_{\rm SAW}$ is the total number of SAW, and $Z_1 (h)$is the
partition function of the one dimensional Ising model in a field $h$.
Using well known results \cite{PGG,desCloizeaux}:
\be
Z_{\rm SAW} \sim \mu^N N^{\gamma - 1}
\ee
where $\gamma$ is a critical exponent, we obtain:
\be
\label{free1}
{F\over N} \le -T \log \mu -T \log (e^{\beta J} \cosh {\beta h}+
\sqrt{e^{2\beta J} \sinh^2 \beta h + e^{-2\beta J}})
\ee
The best estimate in $d=3$ is $\mu \simeq 4.68$ for the cubic lattice
\cite{Gu}.
\item{} At low temperature, the magnetic energy is larger than the
entropy loss due to confinement, and thus the chain collapses. 
The simplest picture is 
that of a  totally magnetized and fully compact system,
resulting in the following bound:
\be
\label{bound2}
Z \ge \exp(N \beta J d + N \beta h) \ Z_{\rm HP}
\ee
where $Z_{\rm HP}$ is the entropy of Hamiltonian paths (HP) on the
lattice, i.e. fully compact SAW on the lattice \footnote{A HP is a
SAW which passes through each point of the lattice exactly once}.
Using a virtually exact upper bound \cite{Orl_Itz_DeD}
to $\ Z_{\rm HP}$, we obtain:
\be
\label{free2}
{F\over N} \le -T \log {2d \over e} - J d - h
\ee
Note that in two dimensions and in zero magnetic field, it is possible 
to write a more accurate bound by using the exact expression $Z_2$ for 
the Onsager partition function of the Ising model:
\be
Z \ge Z_2 \ Z_{\rm HP}
\ee

\end{enumerate}

The free energy bounds of equations (\ref{free1}) and (\ref{free2})
are shown as 
functions of temperature in Figure 1, for $d=3$ and $h=0$.
The true free energy lies below these two curves, and their intersection 
is an indication for a possible first order
transition between the swollen and collapsed phases.

\section{Mean-field theory}
\label{mean_field}
By using a Gaussian transform, 
it is possible to write a field-theoretical representation for the
model (\ref{Z}):
\be
\label{field1}
Z= 2 ^ N \int \prod_{r} d \varphi_r  \exp \left ( -{1 \over 2 \beta J}
\sum_{\{r,r'\}} \varphi_r \Delta ^{-1} _{r,r'} \varphi_{r'} + \log
{\sum_{{\rm SAW} \{r_i\}}
\ \prod_{i=1}^N \cosh ( \varphi_{r_i} + \beta h)} \right )
\ee
As usually, the mean-field theory can be obtained by performing a
saddle-point approximation on equation (\ref{field1}). We assume that the chain
is confined in a volume $V$ with a monomer density $\rho={N\over V}$.
Assuming a translationally 
invariant field $\varphi$, the mean field free energy
per monomer is
\be
\label{meanfield1}
f = {F \over N} =  -T \log2 + {T^2 \over 2 \rho J q} \varphi^2 -T 
\log Z_{\rm SAW} -T \log \cosh (\varphi + \beta h)
\ee
where $q=2d$ is the coordination number of the cubic lattice and
$Z_{\rm SAW}$ is the total number of SAW of $N$ monomers
confined in a volume $V$. Following ref. \cite{Orl_Itz_DeD}, it is
easily seen that:
\be
Z_{\rm SAW} \simeq \left ( {q \over e} \right )^N \exp {\big( - V (1-\rho)
\log (1-\rho)\big)}
\ee
so that
\be
\label{meanfree}
f = -T \log2 + {T^2 \over 2 \rho J q} \varphi^2 -T 
\log {q \over e} + T\ {1-\rho \over \rho}\ \log ( 1-\rho)
-T \log \cosh (\varphi + \beta h)
\ee
This free energy is to be minimized with respect to the field
$\varphi$ and to the volume $V$ occupied by the chain, or equivalently 
to the monomer concentration $\rho$. The mean field equations read:
\bea
T {\varphi^2 \over 2 J q} &=& -\rho - \log (1-\rho) \label{conc}\\
\varphi &=& \beta J \rho q \ \tanh (\varphi + \beta h) \label{mag}
\eea
Note that this free energy holds also for a melt of chains, where
$\rho$ is the total monomer concentration.

This set of coupled equations has a high temperature solution $\rho=0,
\varphi=0$, which describes the swollen phase with no
magnetization and vanishing monomer concentration, 
and a low temperature solution, which describes 
a collapsed phase with 
a finite monomer concentration and magnetization.
More precisely, for magnetic fields $h < h_G$, there is a first order
transition (as a function of temperature) between a swollen and a 
collapsed phase. At higher fields
$h> h_G$, the transition becomes second order (in fact
tricritical). For infinite fields, the magnetization saturates,
and the model becomes equivalent to the ordinary $\Theta$ point 
as studied in many papers \cite{Mei_Lim,Gra_Heg2,TJOW96}. 

Expanding (\ref{conc}) and (\ref{mag}), 
one obtains the equation for the second order line:
\be
\label{crit}
\tanh \beta h = \sqrt{T \over J q}
\ee
Close to this critical line, the concentration varies as:
\be
\rho \sim {1\over2} {1 - \beta J q \tanh^2 \beta h \over (\beta J q) ( 1-
\tanh^2 \beta h ) - {1\over 3}}
\ee
and the magnetization per spin is given by
\be
M \sim  \tanh \beta h
\ee
and remains finite, 
whereas the magnetization per unit volume, given by
\be
m \sim \rho \tanh \beta h
\ee
vanishes.

The phase diagram is shown in Figure 2, with values corresponding
to dimension $d=3$.

The first order and continuous transitions are separated by a
multicritical point, denoted by $G$ on Figure 2. The  corresponding
temperature and field are $T_G=4.5 J$ and $h_G=5.926
J$. At zero magnetic field (point $A$), the transition 
temperature is $T_c=1.886 J$, the critical concentration is
$\rho_c=0.87$ and the critical magnetization per unit volume is $m_c =
0.87$. Note that the magnetic susceptibility $\chi = \partial M /
\partial h$ remains finite along the second order critical line.

The same phase diagram holds for the Heisenberg ferromagnet, with a
multicritical $G$ point at $T_G=2.31 J,\  h_G=5.91 J$. The zero-field
point $A$ is at $T_c=0.844 J$, $\rho_c=0.902$ and $m_c=0.73$.

Using (\ref{mag}) to eliminate
$\varphi$ as a function
of $\rho$, it is possible to
express the free energy (\ref{meanfree}) only as a function of the
monomer concentration $\rho$. This yields the virial expansion of the
free energy. The second virial coefficient vanishes along the
second order line (\ref{crit}), implying that the transition is
$\Theta$-like (i.e. tricritical). At the multicritical $G$ point, both
the second and third virial coefficients vanish, while the fourth
order coefficient is positive. 

\section{Monte Carlo simulations}
\label{MC}

The Monte Carlo method \cite{Bin} used to compute thermodynamic as well as
geometric properties of the magnetic chain relies on
the multiple Markov chain sampling. A detailed description of this
method can be found in \cite{TJOW96,Or}. The implementation we
consider for a magnetic chain on a three dimensional cubic lattice, can
be summarized as follows.  

We start from a single Markov chain at fixed
temperature $T$. The probability $\pi_{{\cal{D}}}(T)$ of a (magnetic) chain
configuration ${\cal{D}}$, is given by the Boltzmann distribution
 $\pi_{{\cal{D}}}(T) \sim e^{- H({\cal{D}})/T}$
with
\begin{equation}
H({\cal{D}}) = -\frac{J}{2}\sum_{i \ne j}
S_i \Delta_{r_i r_j} S_j - h \sum_i S_i 
\end{equation}
where the thermodynamic variables $S_{i}$ and $r_{i}$ are assigned their
${\cal{D}}$-dependent values. This Markov chain is generated by a Metropolis
heath bath sampling based on a hybrid algorithm for chains with pivot
\cite{MS87} a well as local moves \cite{VS61}. Pivot moves are
nonlocal moves that assure the ergodicity of the algorithm; they
operate well in the swollen phase but their efficiency deteriorate
close to the compact phase. In this respect local moves become
essential to speed up the converge of the Markov chain. Finally, in
addition to the moves that deform the chain, an algorithm based on
Glauber dynamics is considered to update the spin configuration along
the chain. For a single Markov chain we typically consider $\sim 10^6$
pivot moves intercalated by $N/4$ local moves and $N$ spin updates. 

The multiple Markov chain algorithm is then implemented on the
hybrid algorithm described above. The idea is to run in parallel
a number $p$ (in this work $p = 20 - 25$) of Markov chains at different
temperatures $T_1 > T_2 > \cdots > T_p$. In practice, this set of
temperatures is such that the configurations at $T_j$ and at $T_{j+1}$
have considerable overlap (implying that $T_{j}$ and $T_{j+1}$  
are close enough). We let the
Markov chains interact by possibly exchanging configurations as
follows. Two neighboring Markov chains (i.e. with temperatures $T_j$
and $T_{j+1}$) are selected at random with uniform probability. A
trial move is an attempt to swap the two current configurations of
these Markov chains. If we denore by $\pi_K(T)$ is the Boltzmann
probability of getting configuration $K$ at temperature $T$, and
${\cal{D}}_j$ and ${\cal{D}}_{j+1}$ are the current states in the
$j$th and $(j+1)$th Markov chain, then we accept the trial move
( i.e. swap ${\cal{D}}_j$ and ${\cal{D}}_{j+1}$) with probability

\begin{equation}
r({\cal{D}}_j,{\cal{D}}_{j+1}) = \min{\left ( 1,
\frac{\pi_{{\cal{D}}_{j+1}}(T_j)\pi_{{\cal{D}}_{j}}(T_{j+1})}
{\pi_{{\cal{D}}_{j}}(T_j)\pi_{{\cal{D}}_{j+1}}(T_{j+1})}\right )}
\end{equation}

The whole process is itself a (composite) Markov chain that is
ergodic, since the underlying Markov chains are themselves
ergodic. 
It turns out that the swapping procedure dramatically decreases the
correlation times within each Markov chain with little cost in CPU
time since, in any case, one is interested in obtaining data
at many temperatures \cite{TJOW96,Or}.

For each multiple Markov chain run we compute estimates, at a discrete
set of temperatues $T$, of quantities such as (i) the average energy
$\langle {\cal E}\rangle$ and specific heat  ${\cal C} = \frac{\langle
{\cal E}^2\rangle - \langle {\cal E}\rangle^2}{T^2}$ of the chain. The
per monomer quantities will be denoted respectively by $E$ and $C$ (ii)
the  average magnetization per monomer $M={1 \over N}\sum_{i}\langle
S_{i}\rangle$ (iii) the susceptibility $\chi={{\partial M} \over
{\partial h}}$. In addition, as a geometric quantity, we consider the
mean squared radius of gyration $\langle R^2\rangle$ of the
chain. From now on, we will set $J=1$, which amounts to give the
values of the field and temperature in units of $J$.

We have done preliminary simulations at high ($T=10$) and low ($T=1$)
temperature. Our results (Figure 3) show that the chain undergoes a
swollen to collapsed phase transition, in broad agreement with the
mean field picture; moreover, the radius of gyration is found to vary
very little with the magnetic field (note that mean field theory
yields $\rho_c =0.87$ at $T_c$ in zero field). Since a full
exploration of the $(h,T)$ plane is difficult, we have restricted this
paper (i) to a detailed study of the $h=0$ transition (ii) to a
qualitative study of some non zero magnetic field transitions. As
mentionned above, the infinite field case corresponds to the usual
$\Theta$ situation: equation (\ref{crit}) then yields
$T_{\theta}=q=6$, whereas the experimental (cubic lattice) value is
$T_{\theta} \simeq 3.7$ \cite{Gra_Heg2}. As expected in the presence
of fluctuations, the mean field parameters (including the location of
point $G$) are not reliable. For small $h$, the first order caracter
of the mean field transition will be seen to improve the situation.  

\subsection{Results for $h=0$: evidence for a first order transition}
\label{h0}
Our results for the specific heat per monomer $C$ and the
susceptibility $\chi$ are respectively given in Figures 4 and 7. The
spiky character of both contrasts with the rounded specific heat of a
usual $\Theta$ point. Indeed, finite size scaling theory \cite{Lam}
predicts that the peak $C_{max}$ of the specific heat behaves, in the
critical region (i.e. for large enough $N$), as 
\begin{equation}
\label{peak}
C_{max} \sim N^{\alpha \over {2-\alpha}}
\end{equation}
 where $\alpha$ is the critical exponent associated with the
temperature divergence of the specific heat. Accordingly, the critical
temperature shifts from its $N=\infty$ value by an amount $\Delta T$
given by 
\begin{equation} 
\label{tempe}
\Delta T \sim N^{-{1 \over {2-\alpha}}}
\end{equation}
At the $\Theta$ point, one has $\alpha=0$, implying a slow
(logarithmic) $N$ dependence of $C_{max}$ and $\Delta T \sim {1 \over
N^{1/2}}$ (up to a logarithmic factor). On the contrary, a thermal 
first order transition corresponds to the value $\alpha=1$, yielding
$C_{max} \sim N$ with a much weaker temperature shift $\Delta T \sim {1 \over
N}$.
These scaling predictions are to be compared with the results of
Figures 5 and 12. The agreement is satisfactory, even though it is not
clear that the largest $N$ value, viz. $N=400$, is already in the
scaling region. Further evidence for a discontinuous zero field
transition comes from Figures 6 and 8, where we show the thermal
evolution of the average magnetization per monomer $M$ and of the
radius of gyration. All these results are consistent with a
first order transition at a critical temperature $T_c \simeq 1.80
\pm.04$, close indeed to the mean field value $T_c^{MF} \simeq 1.88$.

To study the phase coexistence implied by such a transition, we have
studied the probability distributions of the magnetization $M$ and
internal energy $E$ close to the phase transition. Figures 9 and 10,
obtained for $N=300$, suggest that the critical distributions $P(M)$
and $P(E)$ are flat, in 
marked contrast with the usual two peak structure at $T_c$
\cite{Bin_Vol_Deu}. This two peak structure results from the spatial
coexistence of the (bulk) phases along a domain wall (more generally a
$(d-1)$ interface). In the present case, we have coexistence between
phases of different dimensionalities, namely a paramagnetic swollen
phase and a magnetized collapsed phase. It is then clear that the
``interface'' can be reduced to a point, yielding a ``surface''
tension of order one. This in turn explains the flat critical
distributions of Figures 9 and 10. Below the transition, it is
interesting to note that the magnetization quickly saturates: a closer
look at compact chain magnetic conformations shows that the minority
domains are located on the surface of the globule, and become
less and less relevant as $N$ grows (Figure 11). 

\subsection{Tentative studies of the continuous transition in a field}
\label{h1}
As previously mentionned, the multicritical point $G$ will be pushed
downwards from its mean field location. Since the computer search for
a precise determination of this point is very time consuming, we have
adopted the following strategy. We have performed simulations for
small ($h=0.5$ and $h=1$) and large ($h=5$ and $h=10$) magnetic
field. 

The first evidence for a second order transition in large
fields comes from Figure 12, where the specific heat maximum for
$h=10$ behaves very differently from its small field values: for
$h=0.5$ and $h=1$, one apparently gets the same behaviour as with
$h=0$, namely $C_{max} \sim N$. The existence of a second order
transition for $h$ large is corroborated by Figures 14 and 15. For
$h=5$, the critical probability distribution is very different from
its $h=0$ counterpart (Figure 10). A finite size scaling analysis of
the data for the same value of the field yields, in a rather
convincing manner, a second order $\Theta$ like transition at $T_c
\sim 3.4$ (remember that $\lim_{h \to \infty}T_c(h) \simeq 3.7$). We
therefore obtain $h_G <5$. To get a better estimate of $h_G$, we have 
computed the probability distribution $P(E)$ of the internal energy
for $h=0.5$ (Figure 13). It clearly interpolates between Figures 10
and 14, but it is not easy to interpret the data as representative of
a continuous or discontinuous transition. To summarize, we have
presented evidence for a continuous transition for large $h$. The
precise position of the point $G$ is left for future work.
\section{Conclusion}
We have seen that ferromagnetic interactions may drive the collapse of 
a polymer, even in a good solvent. This collapse is very sensitive to
the presence of an external magnetic field. It might be possible to
design new polymeric magnetic materials, for which the collapse
transition is triggered by a magnetic field, at room
temperature.

We have also done preliminary simulations on the two dimensional case
(Ising polymer on a square lattice): for $h=0$, we get a quite abrupt
transition around $T_c \sim 1.18$ (which can be compared to the value
$T_{\theta} \simeq 1.5 $ \cite{Gra_Heg}). Since the critical
dimensions associated with the $\Theta$ point ($\varphi^6$
theory) and the multicritical $G$ point ($\varphi^8$ theory)
are respectively $d_{\Theta}=3$ and $d_{G}={8 \over 3}$, one expects
fluctuations to be important. Further work is needed to elucidate
their influence on the mean field phase diagram.
\newpage
Finally, the present model can be generalized to include \par
\noindent (i) longer range or competing interactions (e.g. ANNNI
models). \par
\noindent (ii) non Ising local variables ($O(n)$ spins,
quadrupoles,...). \par
\noindent (iii) disorder, either in an annealed (BEG-like \cite{BEG})
or in a quenched way \cite{GOP}.  

\begin{figure}
\inseps{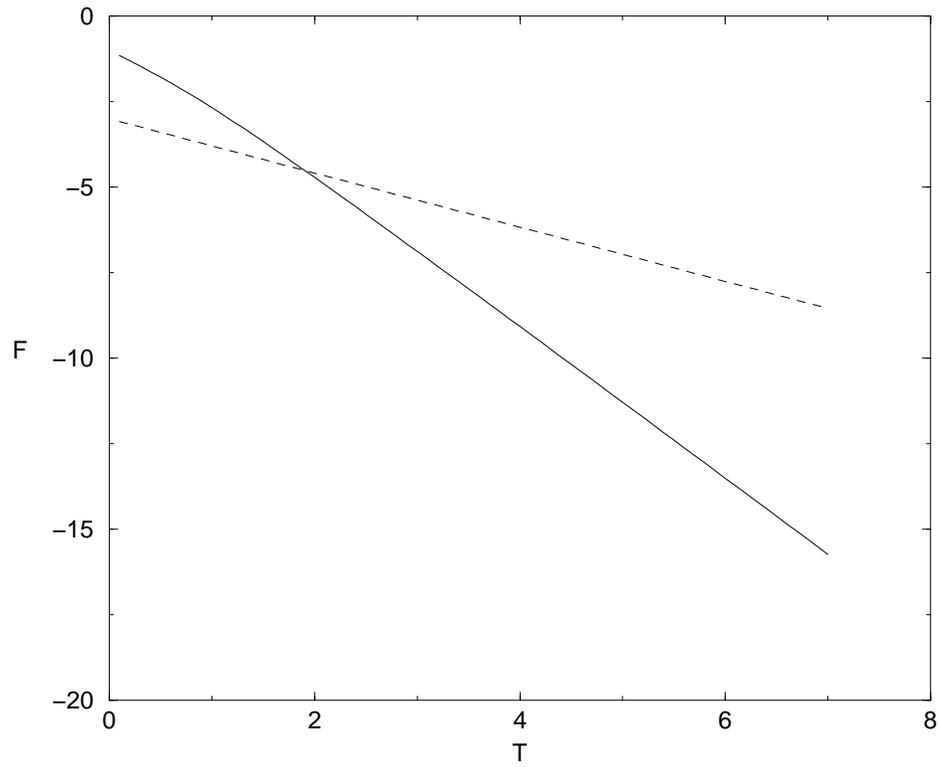}{0.8}
\caption{The high temperature (solid line) and low temperature
(dashed line) free energy bounds for $d=3$ and $h=0$.}
\label{figure1}
\end{figure}

\begin{figure}
\inseps{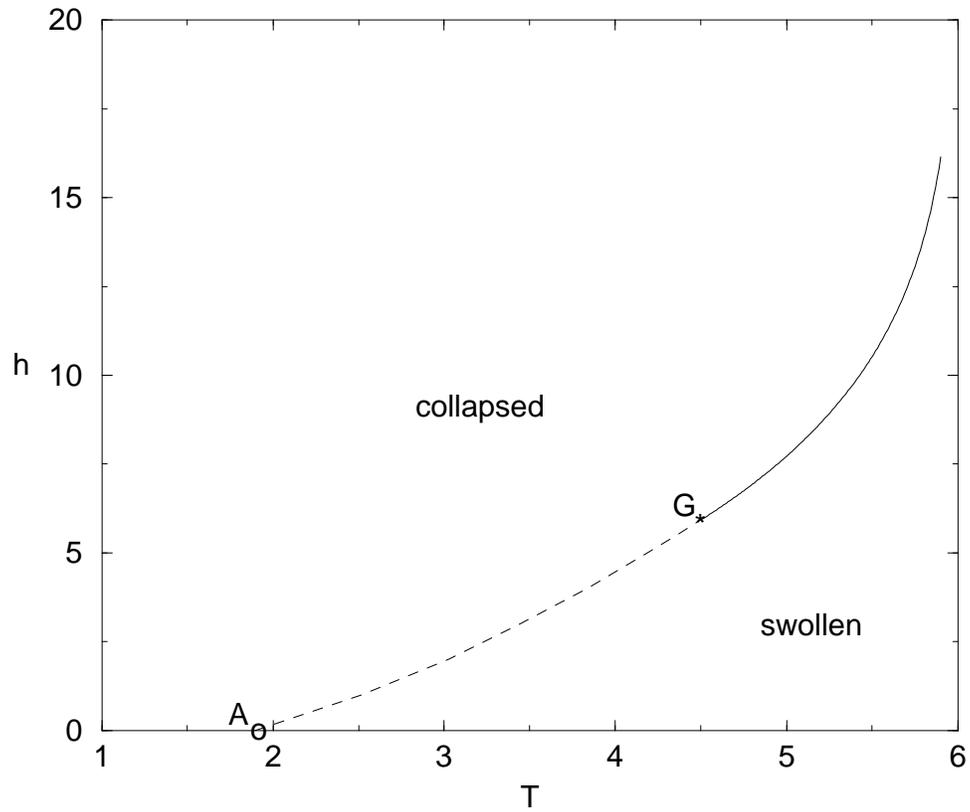}{0.8}
\caption{Mean-field phase diagram in $d=3$ for Ising spins.
The dashed line corresponds to a first order transition, and the
solid line to a second order ($\Theta$-like) transition.
}
\label{figure2}
\end{figure}

\begin{figure}
\inseps{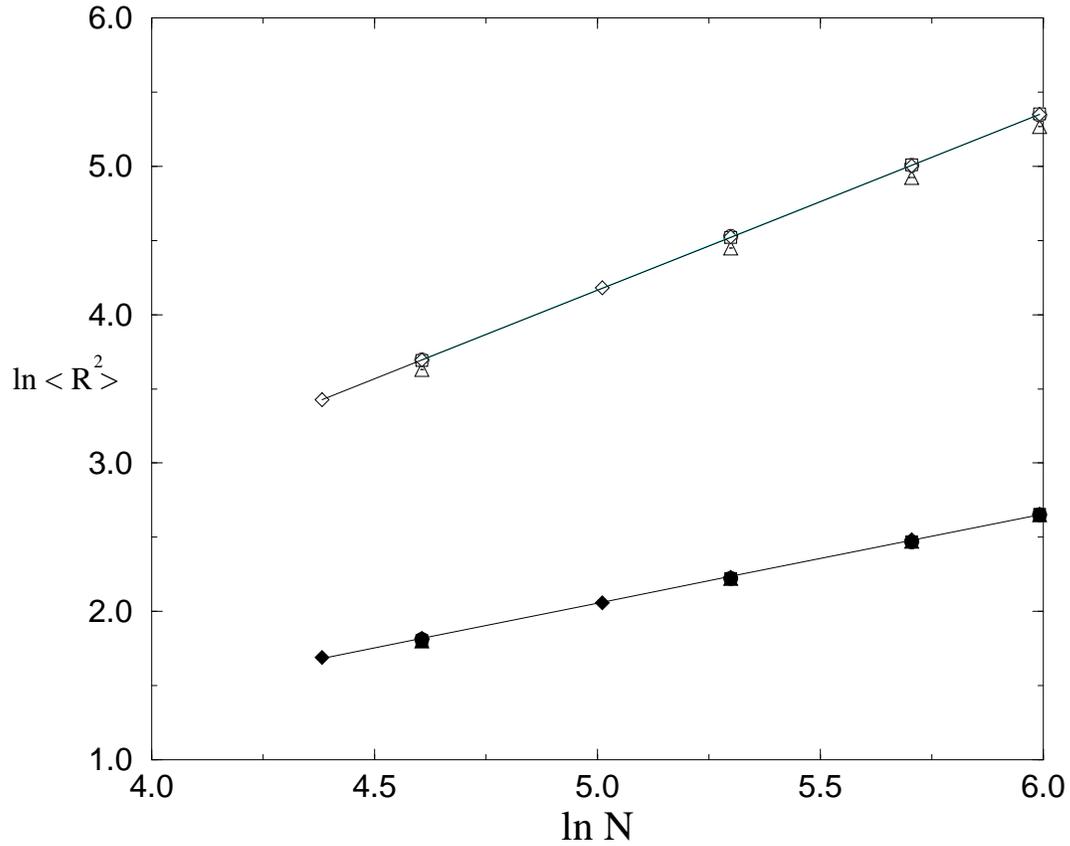}{0.8}
\caption{Log-Log plot of $<R^{2}>$ vs $N$, for $N=80, \ 100, \ 150, \
200, \ 300, \ 400$. The empty and filled symbols
correspond respectively to $T=10$ and $T=1$. Values of the magnetic
field are $h=0 \ (\diamond), \ 0.5 \ (\circ), \ 1 \ (\Box)$ and
$h=10 \ (\triangle)$. The lines have respective slope $2 \nu=1.194
\pm.005$ and $2 \nu=0.63 \pm.04$.} 
\label{figure3}
\end{figure}

\begin{figure}
\inseps{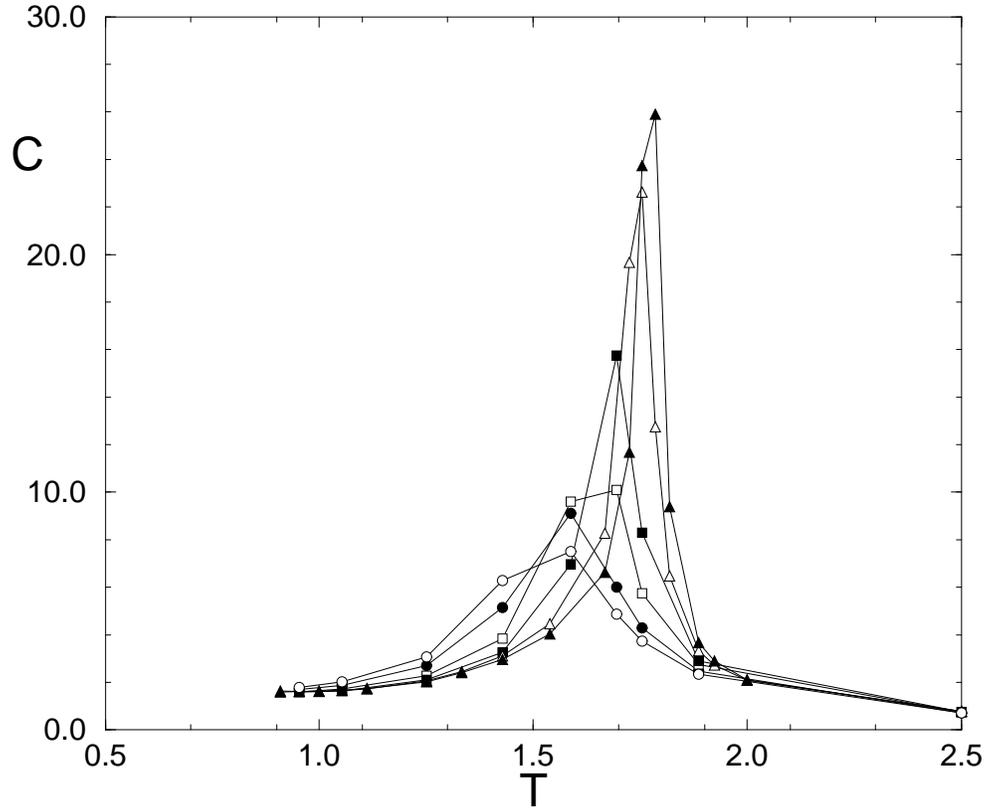}{0.8}
\caption{Specific heat per monomer vs temperature for $h=0$ and $N=80 \
(\circ)$, $\ 100 \ (\bullet)$, $\ 150 \ (\Box)$, $\ 200 \ (\blacksquare)$, $\
300 \ (\triangle)$, $\ 400 \ (\blacktriangle)$. Note the increase of the
peak as  well as its shape, when $N$ increases.} 
\label{figure4}
\end{figure}

\begin{figure}
\inseps{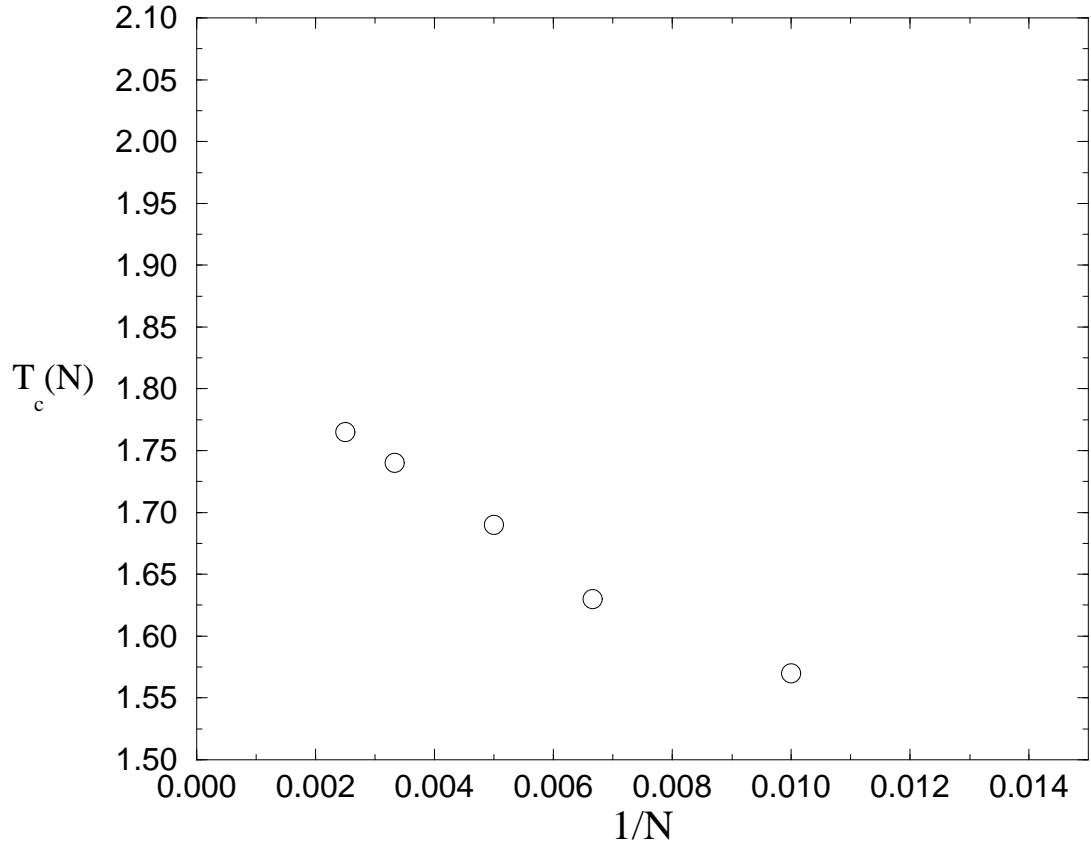}{0.8}
\caption{Temperature location of the specific heat peak vs ${1/N}$,
for $N=100, \ 150, \ 200, \ 300, \ 400$.}   
\label{figure5}
\end{figure}

\begin{figure}
\inseps{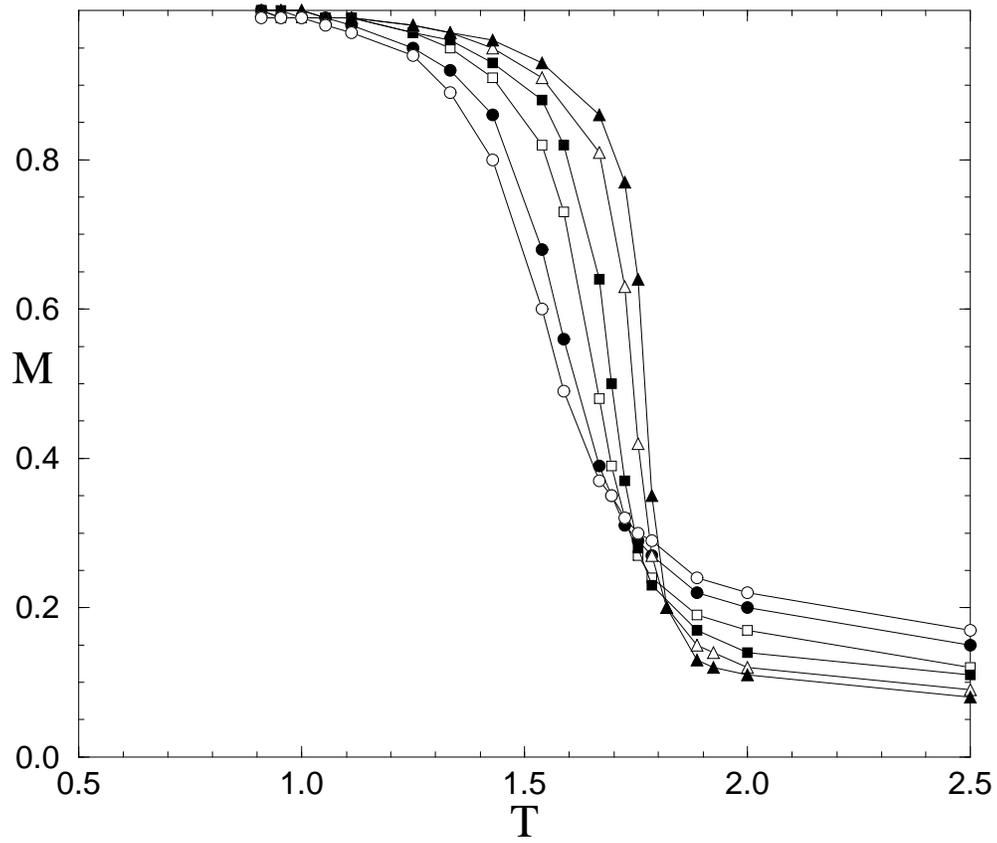}{0.8}
\caption{Magnetization per spin vs temperature for $h=0$ and $N=80 \
(\circ)$, $\ 100 \ (\bullet)$, $\ 150 \ (\Box)$, $\ 200 \ (\blacksquare)$, $\
300 \ (\triangle)$, $\ 400 \ (\blacktriangle)$.}
\label{figure6}
\end{figure}

\begin{figure}
\inseps{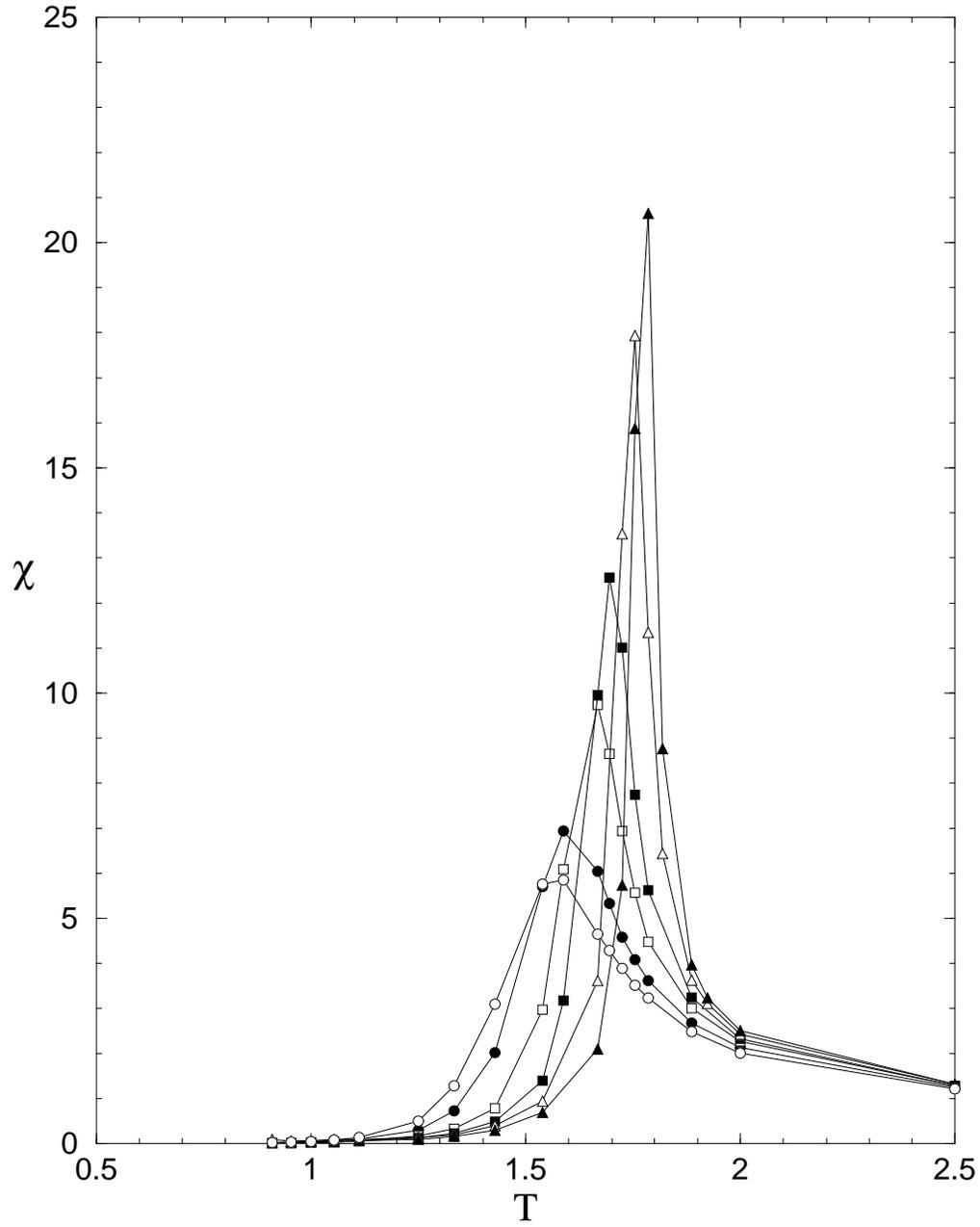}{0.8}
\caption{Susceptibility vs temperature for $h=0$ and $N=80 \
(\circ)$, $\ 100 \ (\bullet)$, $\ 150 \ (\Box)$, $\ 200 \ (\blacksquare)$, $\
300 \ (\triangle)$, $\ 400 \ (\blacktriangle)$.
}
\label{figure7}
\end{figure}

\begin{figure}
\inseps{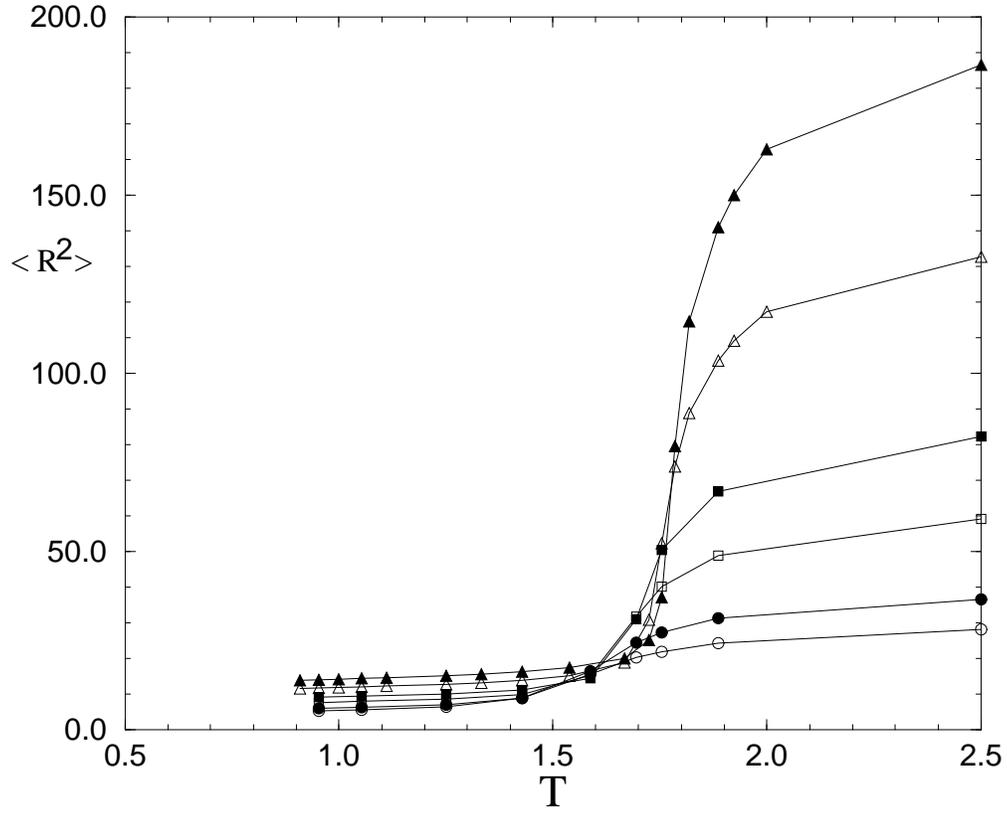}{0.8}
\caption{Squared radius of gyration of the polymer
as a function of  temperature for $h=0$, and
$N=80 \
(\circ)$, $\ 100 \ (\bullet)$, $\ 150 \ (\Box)$, $\ 200 \ (\blacksquare)$, $\
300 \ (\triangle)$, $\ 400 \ (\blacktriangle)$.
}
\label{figure8}
\end{figure}

\begin{figure}
\inseps{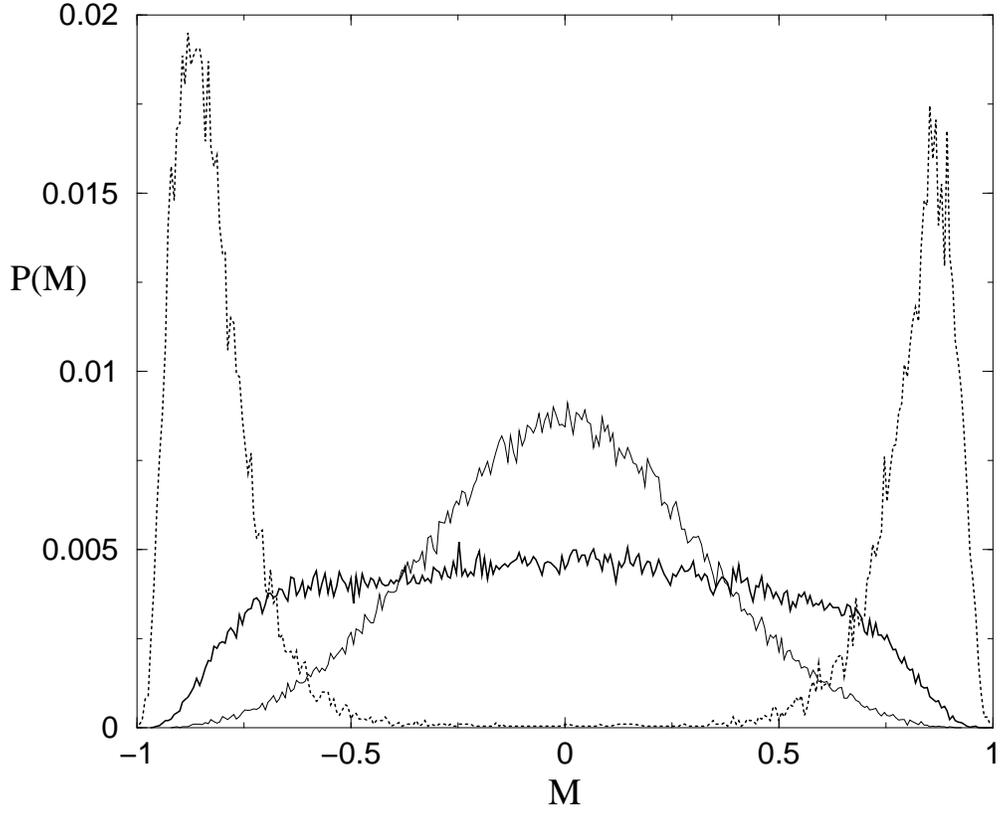}{0.8}
\caption{Probability distribution $P(M)$  of the magnetization per
monomer, for three different temperatures at $h=0$ and $N=300$: $T=
1.67 < T_c(N)$ dotted line, $T = 1.81 > T_c(N)$ solid line, $T = 1.78
\simeq T_c(N)$ thick line.  }    
\label{figure9}
\end{figure}

\begin{figure}
\inseps{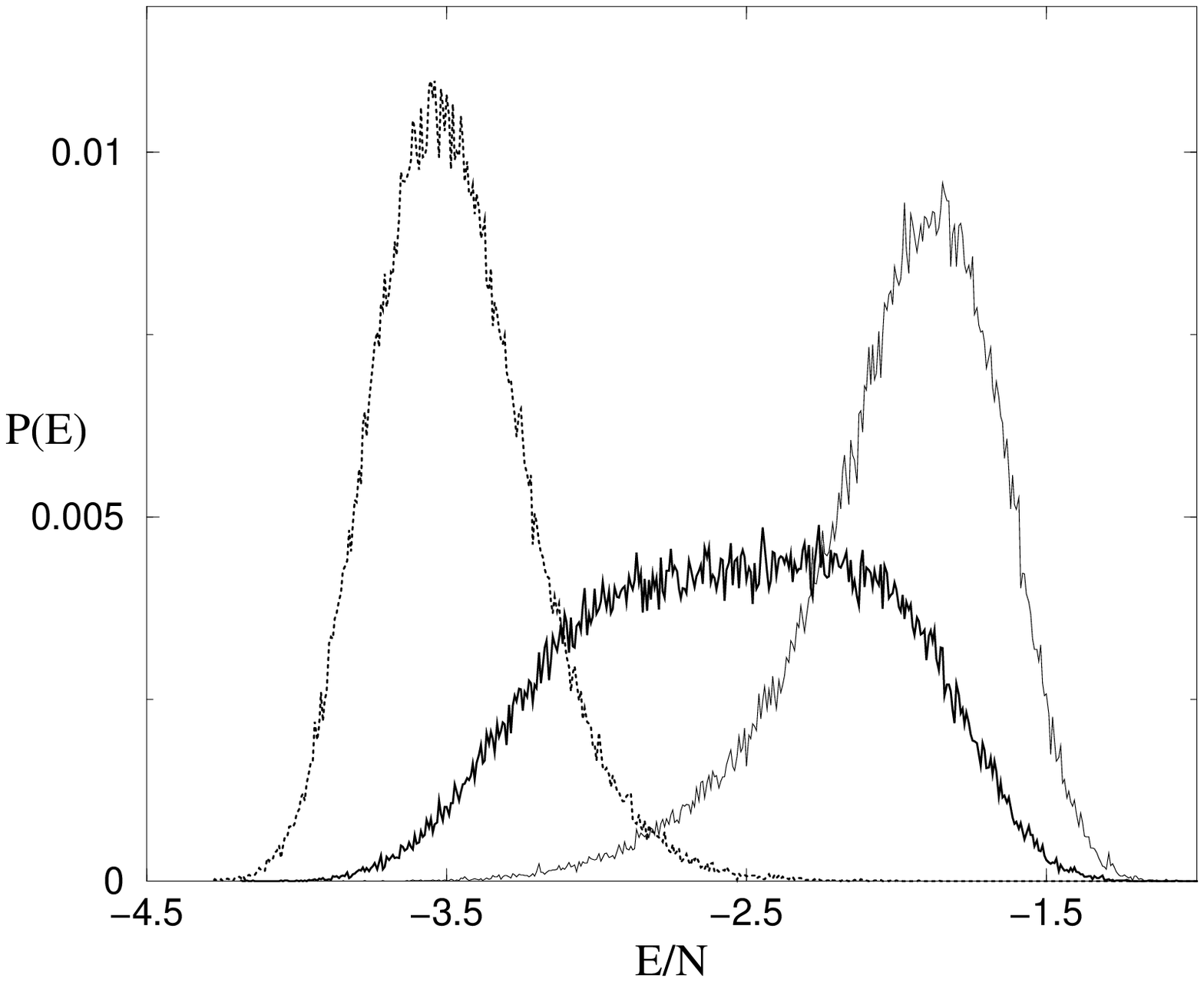}{0.8}
\caption{Probability distribution $P(E)$ of the internal energy per
monomer for three different temperatures at $h=0$ and $N=300$: $T=
1.67 < T_c(N)$ dotted line, $T = 1.81>T_c(N)$ solid line, and $T =
1.78 \simeq T_c(N)$ thick line.  }  
\label{figure10}
\end{figure}

\begin{figure}
\inseps{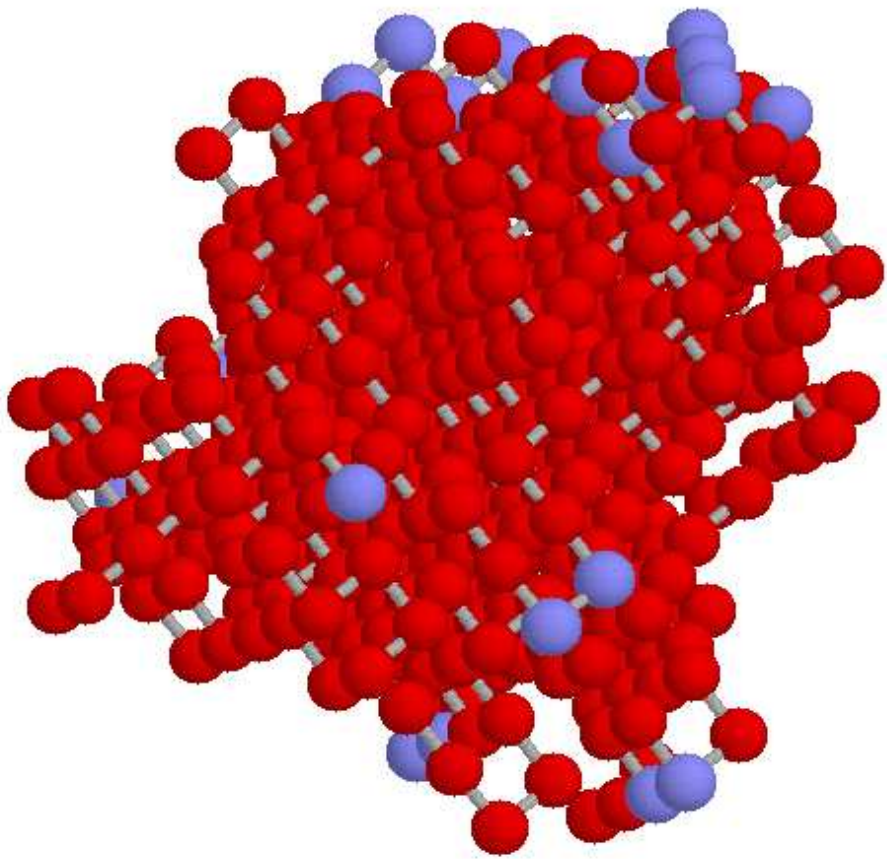}{0.8}
\caption{Typical chain configuration just below the
transition ($T=1.65$), for $h=0$. Red (resp. blue) monomers have up
(resp. down) spins. Note that the down spins are on the surface of the
globule.} 
\label{figure11}
\end{figure}

\begin{figure}
\inseps{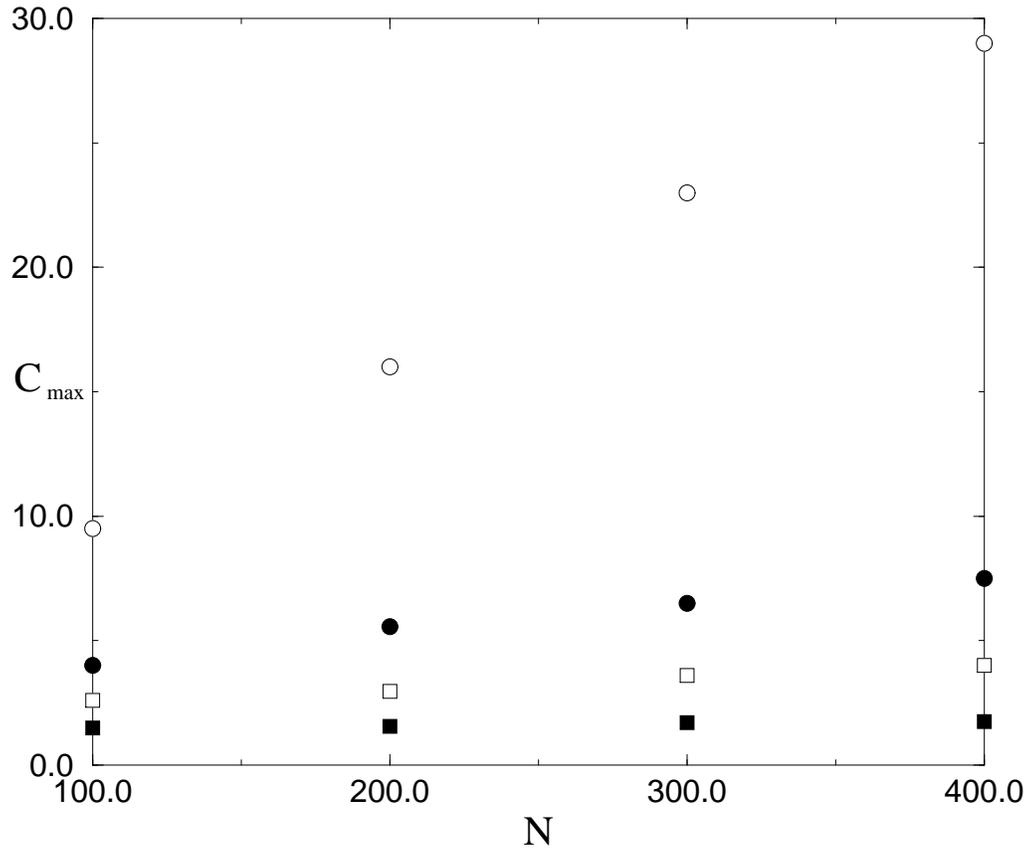}{0.8}
\caption{Peak of the specific heat per monomer vs $N$ for
different values of $h$: $h=0 \ (\circ), \ 0.5 \ (\bullet), \ 1 \
(\Box), \ 10 \ (\blacksquare)$. For $h=10$, the value of $C_{max}$ is
clearly not proportional to $N$.
}
\label{figure12}
\end{figure}

\begin{figure}
\inseps{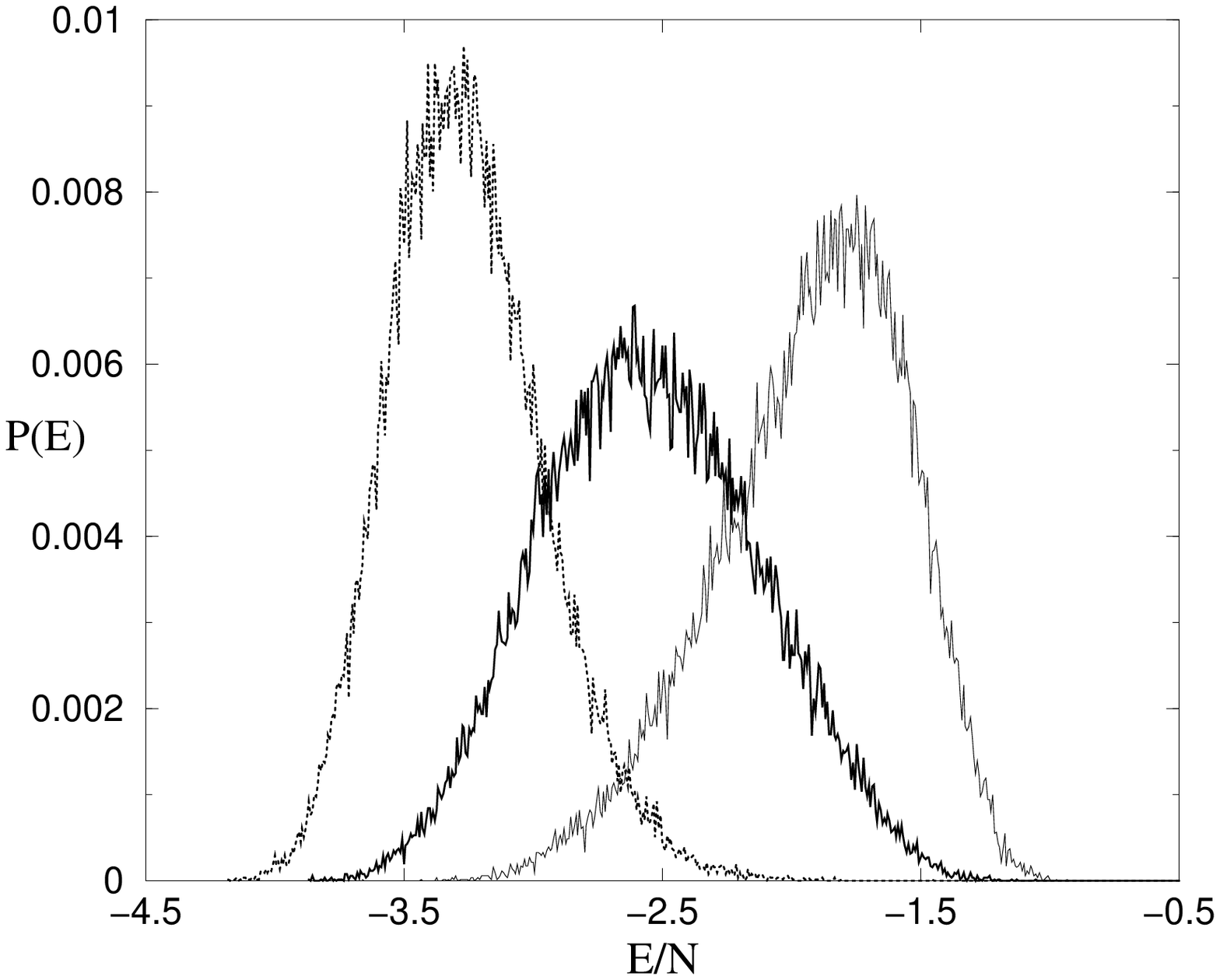}{0.8}
\caption{Probability distribution $P(E)$ of the internal energy per
monomer for three different temperatures at $h=0.5$ and $N=300$: $ T =
1.74 < T_c(N)$  dotted line, $T = 1.89 > T_c(N)$ solid line, $T = 1.81
\simeq T_c(N)$  thick line.  }  
\label{figure13}
\end{figure}

\begin{figure}
\inseps{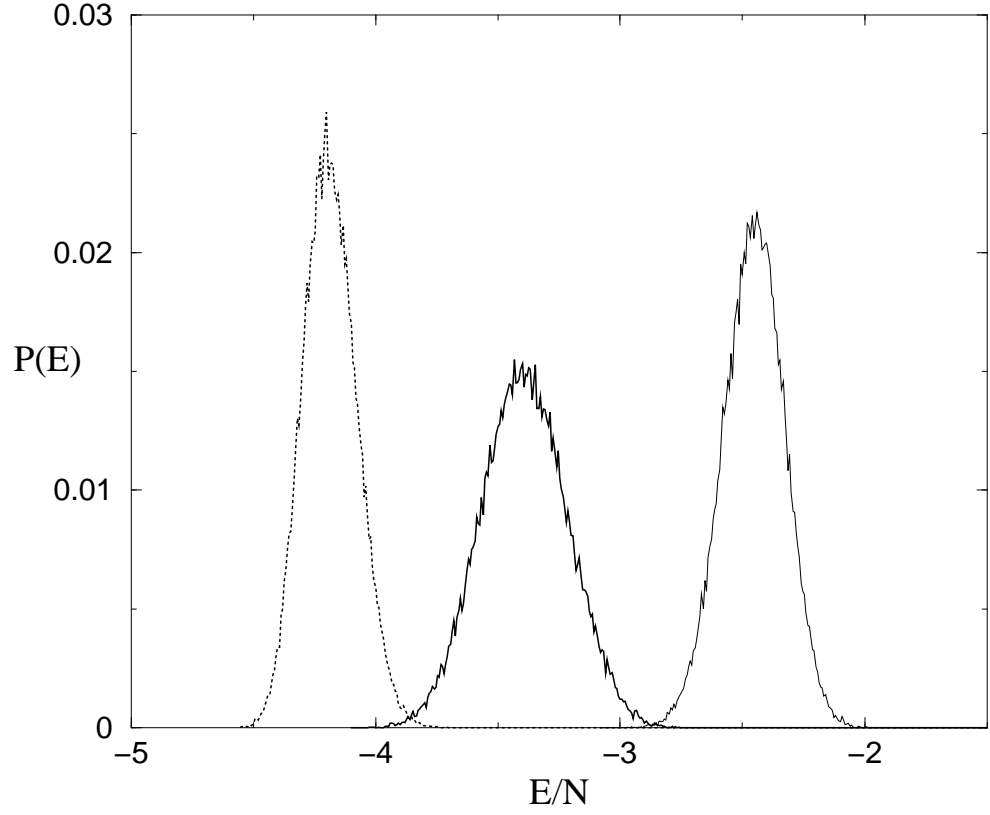}{0.8}
\caption{Probability distribution $P(E)$ of the internal energy per
monomer for three different temperatures at $h=5$ and $N=300$: $T =
1.67 < T_c(N)$ dotted line, $T = 4 >T_c(N)$ solid line, $T =
2.43\simeq T_c(N)$ solid thicker line.  }   
\label{figure14}
\end{figure}

\begin{figure}
\inseps{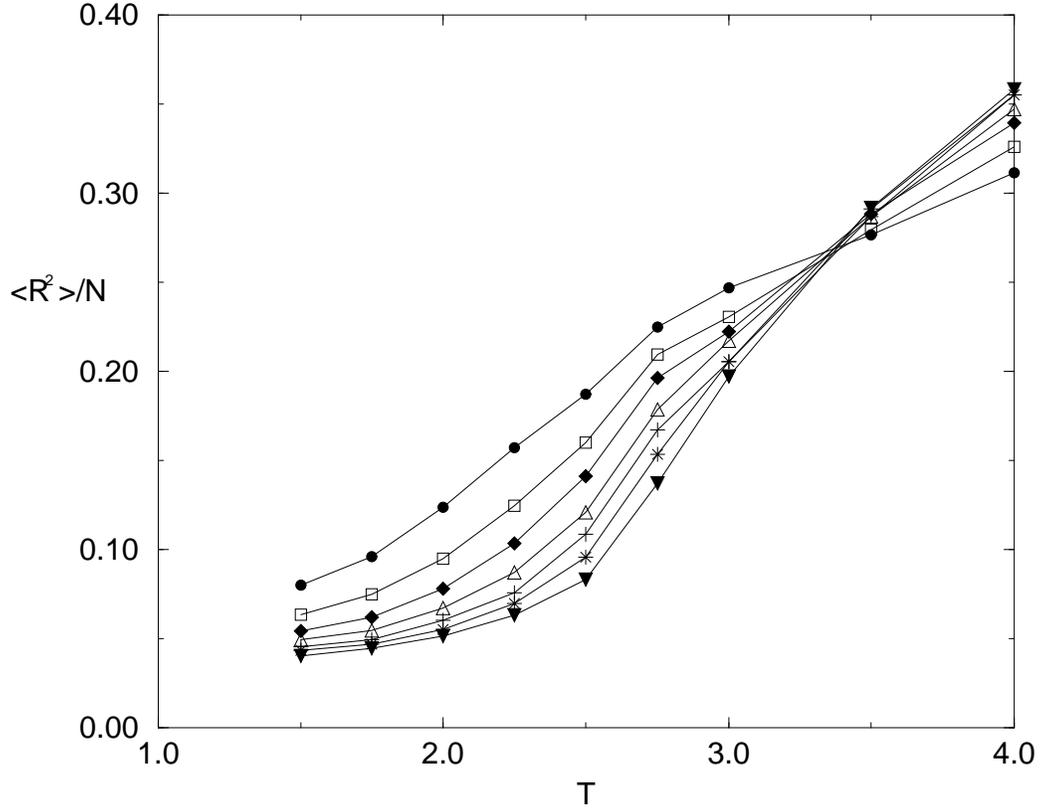}{0.8}
\caption{Search for a second order $\Theta$ like transition for
$h=5$: scaled radius of gyration $\langle{R^{2} \over N}\rangle$ vs 
temperature for $N=100 \ (\bullet)$,$ \ 150 \ (\square)$,$ \ 200 \
(\blacklozenge), \ 250 \ (\triangle), \ 300 \ (+), \ 350 \ (\ast), \
400 \ (\blacktriangledown)$. A crossing occurs for $T \simeq 3.4$.} 
\label{figure15}
\end{figure}



\begin{references}
\bibitem{Flory}
P. Flory, {\it Principles of Polymer Chemistry}, Cornell University
Press (1953).
\bibitem{deGennes}
P.G. de Gennes, J. Phys. (France), {\bf 36}, L55 (1975).
\bibitem{PGO}
E. Pitard, T. Garel and H. Orland, J. Phys. I (France), {\bf 7}, 1201 (1997).
\bibitem{Co_Da} A. Coniglio and M. Daoud, J. Phys. A, {\bf 12}, L259
(1979).  
\bibitem{Ar_Sha}
G.Z. Archontis and E.I. Shakhnovich, Phys. Rev. E, {\bf 49}, 3109
(1994). 
\bibitem{Cha_Ma_Sti} B.K. Chakrabarti, A.C. Maggs and
R.B. Stinchcombe, J. Phys. A, {\bf 18}, L373 (1985). 
\bibitem{Aer_Van} M. Aerstens and C. Vanderzande, J. Phys. A, {\bf 25},
735 (1992).  
%
\bibitem{Mi} J.S. Miller, Adv. Mater., {\bf 6}, 322 (1994).
\bibitem{Ep_Mi} A.J. Epstein and J.S. Miller, Mol. Cryst. Liq. Cryst.,
{\bf 228}, 99 (1993).
\bibitem{PGG} P.G. de Gennes, {\it Scaling concepts in polymer
physics}, Cornell University Press (1979).
\bibitem{desCloizeaux}
J. des Cloizeaux and G. Jannink, {\it Polymers in Solution, their
Modelling and Structure}, Clarendon Press (1990).
\bibitem{Gu} A.J. Guttmann, J. Phys. A, {\bf 22}, 2807 (1989).  
\bibitem{Orl_Itz_DeD}
H. Orland, C. Itzykson and C. De Dominicis, J. Phys. (France), {\bf 
46}, L353 (1985).
\bibitem{Mei_Lim}
H. Meirovitch and H.A. Lim, J. Chem. Phys., {\bf 91}, 2544 (1989).
\bibitem{Gra_Heg2}
P.Grassberger and R. Hegger, J. Chem. Phys., {\bf 102}, 6881 (1995).
\bibitem{TJOW96}
M. C. Tesi,  E. J. Janse van Rensburg , E. Orlandini and S. G. Whittington, 
J. Stat. Phys., {\bf 29}, 2451 (1996).
\bibitem{Bin} K. Binder, {\it The Monte Carlo Method in Condensed
Matter Physics}, Topics in Applied Physics vol 71, Springer (1995).
\bibitem{Or} 
E. Orlandini
{\it  Monte Carlo Study of Polymer Systems by Multiple Markov
Chain Method},to appear in ``Numerical methods for Polymeric Systems", ed. S.
Whittington, IMA Volumes in Mathematics and its Applications,
(Springer Verlag 1998).
\bibitem{MS87}
N. Madras, A. D. Sokal, 
J. Stat. Phys., {\bf 47}, 573 (1987).
\bibitem{VS61}
P. H. Verdier, W. H. Stockmayer 
J. Chem. Phys., {\bf 36}, 227 (1961).
\bibitem{Lam} P.M. Lam, Phys. Rev. B, {\bf 36}, 6988 (1987).
\bibitem{Bin_Vol_Deu} K. Binder, K. Vollmayr, H-P. Deutsch, J.D. Reger
and M. Scheucher, Int. J. Mod. Phys. C, {\bf 3}, 1025 (1992), and
references therein.
\bibitem{Gra_Heg} P. Grassberger and R. Heeger, J. Phys. I (France),
{\bf 5}, 597 (1995).
\bibitem{BEG} M. Blume, V.J. Emery and R.B. Griffiths, Phys. Rev. A
{\bf 4}, 1071 (1971).
\bibitem{GOP}
T. Garel, H. Orland and E. Pitard, ``Protein folding and
heteropolymers'', in {\em Spin Glasses
and Random Fields}, A.P. Young (ed.), World Scientific, Singapore
(1997) p. 387-443.
\end{references}
\end{document}